\documentclass[twoside]{dis09}
\usepackage[latin1]{inputenc}
\usepackage[dvips]{graphicx,epsfig,color}
\usepackage{wrapfig,rotating}
\usepackage{amssymb,amsmath,array}

\newcommand{\simorder}{\raisebox{-4pt}{$\, \stackrel{\textstyle >}{\sim} \,$}}
\newcommand{\simordertwo}{\raisebox{-3pt}{$\, \stackrel{\textstyle <}{\sim} \,$}}

\begin{document}
\title{\mbox{}\\[-5 mm]
Transverse $\Lambda$ Polarization at LHC}

\author{Dani\"{e}l Boer
%
\thanks{Talk presented at the XVII International Workshop on Deep-Inelastic
  Scattering and Related Subjects (DIS 2009), Madrid, April 26-30, 2009}
%
\vspace{.3cm}\\
%
Department of Physics and Astronomy - Vrije Universiteit Amsterdam\\
De Boelelaan 1081 - 1081 HV Amsterdam - The Netherlands\\
}

\maketitle

\begin{abstract}
  Transverse polarization of $\Lambda$ hyperons produced in $p \,
  p$ and $p \, Pb$ collisions is discussed. A factorized
  description in the intermediate to high $\mathbf{p}_T$ region is
  considered that involves transverse momentum and spin dependence in
  the fragmentation process. Consequences and suggestions for
  investigations at LHC are pointed out for the process $p + p \to
  \Lambda^\uparrow + \text{jets} + X$ at midrapidity and $p + p/Pb
  \to \Lambda^\uparrow + X$ in the forward region.
\end{abstract}

\vspace{-3 mm} 

\section{Introduction}
It is well-known since the mid-1970's that $\Lambda$ hyperons produced
in unpolarized $p\, p$ collisions are to a large degree polarized
transversely to the production plane. There have been many
experimental and theoretical investigations aimed at understanding
this striking polarization phenomenon, but no consensus has been
reached about its origin. One of the difficulties in interpreting the
available (mostly fixed target) data is that they are not or only partially in
a region where a factorized description of the cross section is
expected to be applicable. High-energy hadron collider data would be 
very welcome, for instance from RHIC, Tevatron or LHC, but there the
capabilities to measure $\Lambda$ polarization via the self-analyzing 
parity violating decay $\Lambda \to p \, \pi^-$ are typically 
restricted to the midrapidity region, where protons can be identified,
but the degree of transverse polarization $P_\Lambda$ is expected to be 
very small. For symmetry reasons $P_\Lambda=0$ at midrapidity 
in $p\, p$ collisions in the center of mass frame.
Nevertheless, some interesting 
$\Lambda$ polarization studies can be done using the 
process $p + p \to \Lambda^\uparrow +
\text{jets} + X$, where the $\Lambda$ and jets can be in the midrapidity
region without paying a suppression penalty. It is
especially of interest at LHC, where the asymmetry expressions may take
a particularly simple form depending on the importance of gluons. 
Also, transverse $\Lambda$ polarization 
measurements at forward rapidities at LHC will be discussed below, as
it offers a promising way of extracting the $x$ dependence of the
saturation scale. These two suggestions will hopefully
enhance the interest in $\Lambda$ polarization measurements at high
energy colliders, at LHC in particular.

\vspace{-2 mm}  

\section{Transverse $\Lambda$ polarization in unpolarized collisions}
Large asymmetries have been observed in $p + p \rightarrow
\Lambda^{\uparrow} + X$ \cite{Lesnik:1975my}. 
The main features of the asymmetry are:
$|P_{\Lambda}|$ grows with $x_F^{}$ and $p_T^{} 
\ (\simordertwo 1\, \text{GeV}/c$); for $p_T^{} \simorder 1\, \text{GeV}/c$ 
it becomes flat (measured up to $4$ GeV/c); no $\sqrt{s}$ dependence
has been seen. For a comprehensive review of data cf.\ Ref.\  
\cite{Panagiotou:1989sv}.
Many QCD-inspired models have been proposed to explain the transverse
$\Lambda$ polarization data. 
Most models give qualitative descriptions of the data for $p_T^{}
\simordertwo 1-2\, \text{GeV}/c$. However, $P_\Lambda$ stays large 
at least until the highest measured $p_T \sim 4$ GeV/$c$. For
sufficiently large 
$p_T^{}$ perturbative QCD and collinear factorization should become
applicable. 
Consider for example the $q g \to q g$ subprocess contribution to
the $p \, p \to \Lambda \, X$ cross section in collinear
factorization. It is of the form:
$\sigma \sim {q(x_1)}  \otimes {g(x_2)}  \otimes
\hat{\sigma}_{q g \to q g} \otimes {D_{\Lambda/q}(z)}$, where
${q(x_1)}$ is the quark density in proton 1, ${g(x_2)}$ is the gluon
density in proton 2, and 
${D_{\Lambda/q}(z)}$ is the $\Lambda$ fragmentation function (FF). In a
similar way the transverse polarization should be of the form:
$P_{\Lambda} \sim q(x_1)  \otimes g(x_2) \otimes
\hat{\sigma}_{q g \to q g} \otimes {?}$, involving the unpolarized
parton densities and the unpolarized hard partonic subprocess. 
The latter because $\Lambda$ polarization created
in the helicity conserving hard partonic scattering is very small, 
$P_\Lambda \sim \alpha_s m_q/\sqrt{\hat s}$ \cite{Kane:1978nd}.
The question mark indicates
that at leading twist there is no collinear fragmentation function
describing $q \to \Lambda^\uparrow X$ for symmetry reasons. In
collinear factorization $P_{\Lambda}$ is necessarily power
suppressed. 
Dropping the demand of {\em collinear\/} factorization, 
does allow for a leading twist solution: the function 
\begin{figure}
\begin{center}
\psfig{file=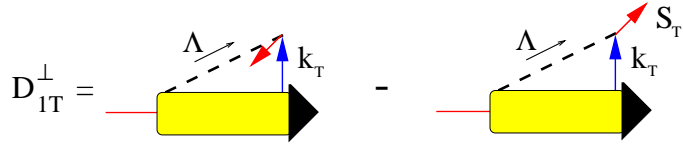,width=3.5in}
\caption{Nonzero $D_{1T}^\perp$ for $\Lambda$'s means that
the transverse polarization $S_T$
  of $\Lambda$'s (with momentum $K_\Lambda \approx z q+k_T$) arising
  from a fragmenting unpolarized quark (with momentum $q$) is
  nonzero. It is a $k_T$-odd, chiral-even TMD fragmentation function.}
\label{D1Tperp}
\end{center}
\end{figure}
$D_{1T}^\perp(z,{\mathbf{k}_T})$ \cite{Mulders:1995dh} for $\Lambda$'s, 
depicted in Fig.\ \ref{D1Tperp}. It describes a nonperturbative 
{$\mathbf{k}_T \times \mathbf{S}_T$} dependence in the fragmentation
process, which is allowed by the symmetries (parity and time
reversal). As the $\Lambda$ polarization arises in the fragmentation of 
an {{\em unpolarized}} {quark}, the descriptive name 
``polarizing fragmentation function'' was suggested for it
\cite{Anselmino:2000vs}. Currently ${D_{1T}^\perp}$ is expected 
to be {universal}, despite its potential {color flow dependence}
\cite{Metz:2002iz}.
${D_{1T}^\perp}$ has been extracted from fixed target 
$p + p/Be \to \Lambda^{\uparrow}/\bar\Lambda^{\uparrow} + X$ data
\cite{Anselmino:2000vs}. Reasonable 
functions are obtained: $D_{1T}^\perp$ has {opposite signs for $u/d$ versus 
$s$ quarks, and the latter is larger. This leads to cancellations
in order that {$P_{\bar \Lambda} \approx 0$}.
This extraction has been done under the restriction of {$p_{T} > 1$
  GeV/$c$} to exclude the soft regime, but also to retain
sufficient data to make a fit to. Whether this restriction 
is sufficiently strict to ensure the validity of the description 
is a matter of concern, due to the large $K$ factors required to
obtain a cross section description.

\vspace{-2 mm} 

\section{Jet-$\Lambda^\uparrow$~production}

The validity of the factorized description depends on whether a
proper cross section description can be obtained. 
This requires data at higher $\sqrt{s}$ and $p_T$, but not necessarily
also at large $x_F$ if one goes beyond $p \, p \rightarrow
\Lambda^{\uparrow} \, X$. 
If the origin of the transverse $\Lambda$ polarization
is indeed due to polarizing fragmentation, then another related asymmetry
could be observed that does not need to vanish at $\eta_\Lambda=0$, 
namely in the process {$p \, p \to \left(
\Lambda^\uparrow \text{jet}\right) \, \text{jet} \, X$} \cite{Boer:2007nh}.
The suggestion is to select two-jet events and to measure
the jet momenta $K_j^{}$ and $K_{j'}^{}$ (with $K_{j}^{}\cdot
K_{j'}^{} = {\cal O}(\hat{s})$), in addition to the momentum
$K_\Lambda^{}$ and polarization $S_\Lambda^{}$ of the $\Lambda$ that
is part of either of the two jets. A single spin asymmetry proportional to
$\epsilon_{\mu\nu\alpha\beta} K_j^{\mu} K_{j'}^{\nu}
K_\Lambda^{\alpha} S_\Lambda^\beta$ can then arise, which is neither
power suppressed, nor needs to be zero at midrapidity. 
In the center of mass frame of the two jets the asymmetry is of the form:
\begin{equation}
\text{SSA}
=\frac{d\sigma({+}\boldsymbol S_\Lambda)\,
{-}\,d\sigma({-}\boldsymbol S_\Lambda)}
{d\sigma({+}\boldsymbol S_\Lambda)\,
{+}\,d\sigma({-}\boldsymbol S_\Lambda)}
=\frac{\boldsymbol {\hat K}{}_j{\cdot}
({\boldsymbol K_\Lambda{\times}\boldsymbol S_\Lambda})}{
z\,M_\Lambda}\,{\frac{d\sigma_T}{d\sigma_U}}\ 
\end{equation}
The analyzing power {$d\sigma_T/d\sigma_U$} of the asymmetry 
depends on $D_{1T}^\perp$. This new
$\Lambda$+jets observable could allow for a more trustworthy extraction
of $D_{1T}^\perp$ (for both quarks and gluons) and subsequent
predictions, for instance for
semi-inclusive DIS \cite{Anselmino:2001js}.

At LHC (and at RHIC) this process $p \, p \to \left(
\Lambda^\uparrow \text{jet}\right) \, \text{jet} \, X$ can be studied.
For instance, ALICE with its excellent PID capabilities can measure 
$\Lambda$'s over a wide $p_T$ range (for example, in a typical
yearly heavy ion collision run at least up to 16 GeV/$c$). 
The ALICE rapidity coverage is ${-}0.9\,{\leq}\,\eta\,{\leq}\,{+}0.9$.
Jets can be reconstructed above 30 GeV (up to 250 GeV in a typical
yearly run). If the jet rapidities ($\eta_{j,j'}$) are {in this
kinematic region and if gluon fragmentation is at least
as important as quark fragmentation for both unpolarized and polarized
$\Lambda$ production, then the process
is dominated by gluon-gluon ($gg{\rightarrow}gg$) scattering}\footnote{Unlike in Ref.\ \cite{Boer:2007nh}, here it will be assumed 
that universality of $D_{1T}^\perp$ holds throughout. Furthermore, 
chiral-odd contributions \cite{Koike} will not be considered here.}:
\begin{equation}
\frac{d\sigma_T}{d\sigma_U}
\approx
{\frac{D_{1T}^{\perp\,g}(z{,}K_{\Lambda\,T}^2)}{D_1^g(z{,}K_{\Lambda\,T}^2)}}.
\end{equation}
Because no model or fit for $D_{1T}^{\perp\,g}$ is available yet, no
predictions can be made in this case. 

If it happens that {$D_{1T}^{\perp\, g}\,{\ll}\,D_{1T}^{\perp\, q}$}, then
one can use the extracted $D_{1T}^{\perp\,q}$ to obtain an
estimate. When gluons still dominate in the denominator, one finds 
for $\eta_{j'}{\approx}\,{-}\eta_j$ ($x_1 \approx x_2$)
\begin{equation}
\frac{d\sigma_T}{d\sigma_U}
\approx {\left[\,b(y){+}b(1{-}y)\,\right]
\frac{\sum_q f_1^q(x_1)D_{1T}^{\perp\,q}(z{,}K_{\Lambda\,T}^2)}
{f_1^g(x_1) D_1^g(z{,}K_{\Lambda\,T}^2)}},
\end{equation}
where ${y\,{=}\,(e^{2\eta_j}{+}1)^{-1}}$,  
$x_1\,{\approx}\,x_\perp/2\sqrt{y(1{-}y)}$, and ${b(y)\, 
=\, d\hat\sigma_{qg\rightarrow qg}/d\hat\sigma_{gg\rightarrow gg}}$.
In the considered rapidity interval the prefactor {$b(y){+}b(1{-}y) \
(\approx 0.4)$} is almost $y$ independent.

In practice, it may be that $D_1^g$ is considerably smaller than
$D_1^q$. In that case the $qg\to qg$ subprocess needs to be taken into
account in the denominator of the asymmetry too (not done 
\begin{wrapfigure}{l}{0.6\columnwidth}
\centerline{\includegraphics[width=0.65\columnwidth]{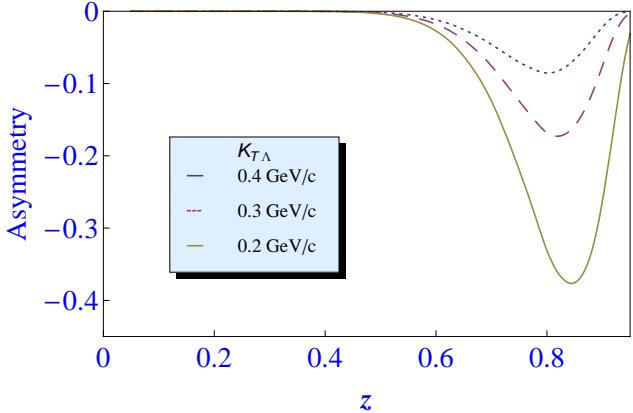}}
\caption{The asymmetry $d\sigma_T/d\sigma_U$ for $\eta_{j},\eta_{j'} = 0$ and 
$|K_{\perp\, j}|, |K_{\perp\, j'}| = 70 $ GeV, using DSV FFs.}
\label{Asymm}
\end{wrapfigure}
in \cite{url}). Here we will use the DSV fragmentation functions of Ref.  
\cite{deFlorian:1997zj} which indeed 
has $D_1^g \ll D_1^q$ at larger $z$. 
The asymmetry 
is given in Fig.\ \ref{Asymm} for three different values of the
$\Lambda$ momentum component transverse to the jet direction.
For very low values of this $K_{\Lambda T}$ the asymmetry at large
$z$ exceeds -1, which is unphysical. It signals a problem with the
$k_T$-dependence of the function $D_{1T}^{\perp\,q}$ for which it was 
not properly taken into
account that for each value of $K_{\Lambda T}$ the positivity bound
has to be satisfied \cite{Anselmino:2001js}. 
Nevertheless, it may be expected that the result
at least has the generic
shape for negligible $D_{1T}^{\perp\,g}$. 
The asymmetry is quite sensitive to the
cancellation between $u/d$ and $s$ contributions, like in SIDIS
\cite{Anselmino:2001js}, and can even flip its overall sign depending on the
amount of $SU(3)$ breaking in the unpolarized fragmentation functions (DSV
assumes $SU(3)$ symmetry). 
More reliable estimates are not possible at this stage.
ALICE would have most data in the region $z <0.5$ and would therefore 
provide valuable information on gluonic contributions to 
$D_{1T}^\perp$. 

\section{$\Lambda^\uparrow$ at forward rapidities and gluon saturation}

$\Lambda$ polarization is also very interesting in 
{$p \, A$ reactions at very high $\sqrt{s}$, large $A$ and
  $\eta$}. In that kinematic regime of {small $x$}, 
{saturation of the gluon density is expected}. 
The process $p \, A \to \Lambda^\uparrow \, X$ 
is sensitive to saturation and could help
to determine properties of this phenomenon. 
{None of the existing
  data is in the saturation regime}. 
At high energy colliders, such as RHIC and LHC, protons often 
cannot be identified in the
forward direction, which hampers the measurement of $\Lambda$ polarization. 
Although relatively forward $\Lambda$'s ($y=2.75$) in $d \, Au$ collisions have
been identified through event topology \cite{Abelev:2007cc}, 
it is not clear whether
the polarization can be reconstructed in this way too, despite the 
self-analyzing decay property. An alternative may be to use neutral decays 
$\Lambda \to n\, \pi^0$} (50\% less frequent than $p
\pi^-$). Despite being a very challenging measurement, 
$\Lambda$ polarization at forward rapidities offers a unique direct probe of
gluon saturation in both $p\, p$ and $p\, Pb$ collisions at LHC.
{The saturation scale $Q_s$ and even
its evolution with $x$ could be probed in this way}
\cite{Boer:2002ij,Boer:2008ze}.

\begin{figure}[htb]
\begin{center}
\psfig{file=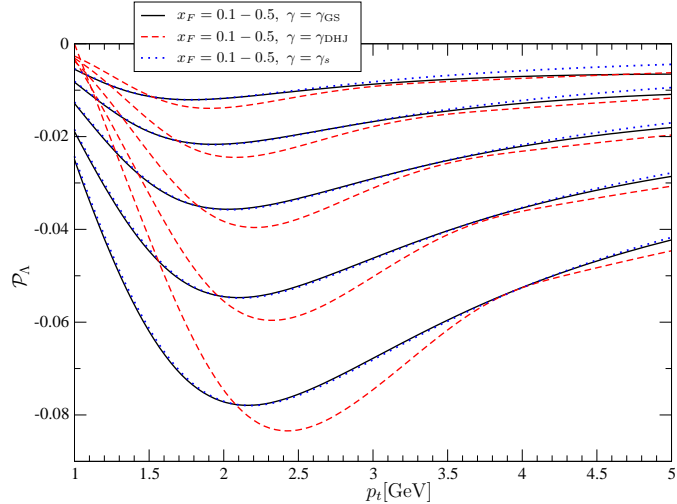,width=3.5in}
\caption{Predictions of $\Lambda$ polarization in 
$p + Pb \to \Lambda^\uparrow + X$ at $\sqrt{s}=8.8$ TeV \cite{Boer:2008ze}.}
\label{AUplot}
\end{center}
\end{figure}

The cross section of forward hadron production in the (near-)saturation
regime is schematically of the form:  
{pdf} $\otimes$ {dipole cross section} $\otimes$ {FF} \cite{Dumitru:2002qt}.
{Since $D_{1T}^\perp$ is $\mathbf{k}_T$-odd, it essentially probes the 
derivative of the dipole cross section}.
At transverse momenta of order $Q_s$ the dipole cross section changes
much, which thus leads to {a $Q_s$-dependent peak in (strictly speaking minus) 
the $\Lambda$ polarization}. This was first demonstrated in Ref.\
\cite{Boer:2002ij} for the 
McLerran-Venugopalan model \cite{McLerran:1993ni}.
{In this model $Q_s$ is a constant, leading to a peak that is $x_F$
independent}. More realistically the saturation scale is expected to
change with the small-$x$ values probed: ${Q_s^2(x) \propto x^{-\lambda}}$.
Models that incorporate this are for instance: the well-known {GBW
  model} \cite{GolecBiernat:1998js}, which describes well 
{small-$x$ DIS data}; the {DHJ model} \cite{Dumitru:2005gt}, which 
describes well {forward $d \, Au \to \pi X$ RHIC data}; and, the {GS
  model} \cite{Boer:2007ug}, which describes well both types of
data. In all three models $\lambda = 0.3$, as indicated by the DIS data. 
Here we will restrict to the latter two
models. Although they have
considerable differences (the GS model is geometrically scaling,
the DHJ model is not; the GS model leads to more steeply falling $p_T$ 
spectra of produced hadrons) \cite{Boer:2007ug}, 
both the {DHJ and GS models lead
to the same conclusion about the peak
of the $\Lambda$ polarization:} {its $x_F$ dependence is to very good 
approximation the $x$ dependence of $Q_s$!} This is shown in Fig.\
\ref{AUplot} for $p \, Pb$ collisions at LHC (the solid line is the 
GS model prediction, the dashed line the DHJ one).  
In $p\, p$ collisions at LHC $Q_s$ and hence the $\mathbf{p}_T$
position of the 
peak is slightly lower. At RHIC unfortunately 
the peak is likely situated below 1 GeV/$c$, where
the formalism cannot be trusted. 
Therefore, 
{$\Lambda$ polarization studies at LHC could prove most interesting.}

\section*{Acknowledgments}

First of all, I thank my various collaborators on this topic. Second, I thank 
Gerry Bunce and Carl Gagliardi for useful discussions. 
Part of this research was funded by the ``Stichting voor
Fundamenteel Onderzoek der Materie (FOM)'', which is financially supported
by the ``Nederlandse Organisatie voor Wetenschappelijk Onderzoek (NWO)''.


\begin{footnotesize}

\end{footnotesize}


\end{document}